
\input phyzzx.tex
\def\a{\alpha}
\def\b{\beta}
\def\g{\gamma}

\def\d{\delta}

\def\e{\epsilon}

\def\l{\lambda}

\def\o{\omega}
\def\O{\Omega}
\def\s{\sigma}

\def\pa{\partial}
\def\na{\nabla}

\def\ov{\overline}
%

\def\pl#1{{\it Phys. Lett.} {\bf #1B}}

\def\prd#1{{\it Phys. Rev.} {\bf D#1}}

\def\np#1{{\it Nucl. Phys.} {\bf B#1}}

\def\mpl#1{{\it Mod. Phys. Lett.} {\bf A#1}}

%

\REF\mtw{C.W. Misner, K. S. Thorne, and J.A. Wheeler,{\it Gravitation} W. H.
Freeman  (1973).}
\REF\beck{J.D. Bekenstein, \prd{5} (1972) 1239 ,J. Hartle in {\it Magic without
magic: John Archibald Wheeler} Freeman (1972),C. Teitelboim, \prd{5} 2941
(1972).}

\REF\adler{S.L. Adler, and R.P. Pearson, \prd{18},(1978) 2798.}
\REF\gib{G. Gibbons and K. Maeda \np{298} (1988) 741.}

\REF\lee{K. Lee and E. Weinberg Columbia preprint (april 1991) CU-TP-515.}
\REF\andy{D. Garfinkle, G.T. Horowitz, and A. Strominger, \prd{43} (1991)
3140.}
\REF\al{A. Shapere, S. Trivedi and F. Wilczek IAS preprint (1991)
IASSNS-HEP-91.}
\REF\wit{E. Witten, \prd{44} (1991) 314.}
\REF\man{G. Mandal, A. Sengupta, and S. Wadia, \mpl{A6} (1991) 1685.}
\REF\dea{S.P. de Alwis and J. Lykken \pl{269} (1991) 264.}
\REF\sch{J. Schwarz, Caltech preprint (1991) CALT-68-1728.}
\REF\ellis{J. Ellis, N. Mavromatos and D. Nanopoulos, CERN  preprint (1991)
CERN-TH-6147/91.}
\REF\call{C. Callan, D. Friedan,
E. Martinec, and M. Perry, \np{262} (1985) 593.\hfil\break
E.S. Fradkin and A. Tseytlin,
\pl{158} (1985) 316; \np{261} (1986) 413.}\hfil\break
\REF\sda {S.P. de Alwis, \pl{168} (1986) 59, and unpublished notes.}
\REF\ramy{R. Brustein and S.P. de Alwis preprint (1991) COLO-HEP-259,UTTG-25-91
\pl{} to be published.}
\REF\dj{S.R. Das and A. Jevicki, Mod. Phys. Lett. A5 (1990) 1639.}
\REF\jev{K. Demeterfi, A. Jevicki, and J.P. Rodrigues,
Nucl. Phys.  {\bf B362} (1991) 173; {\bf B365} (1991) 499.}
\REF\cole{S. Coleman {\it Aspects of Symmetry} Cambridge 1989.}
\REF\ss{N. Seiberg and S. Shenker Rutgers preprint (1992) RU-91-53.}
\REF\wat{R. Brustein, D. Nemeschansky and S. Yankielowicz, Nucl. Phys. B301
(1988) 224; T. Itoi and Y. Watabiki, Phys. Lett. B198 (1987) 486; A. A.
Tseytlin, Johns Hopkins preprint (1991).}
\REF\hh{G.W. Gibbons and S. W. Hawking, \prd{15}  (1977) 2752; G. W. Gibbbons
and M.
J. Perry, Proc. Roy. Soc. London, {\bf A358} (1978) 467.}
\pubnum {COLO-HEP-267}
\date={January, 1992}
\titlepage
\vglue .2in
\centerline{\bf Comments on No-Hair Theorems and Stabilty of Blackholes}
\author{ S.P. de Alwis\foot{dealwis@gopika.colorado.edu}}
\address{Dept. of Physics, Box 390,\break
University of Colorado,\break Boulder, CO
80309}
\vglue .2in
\centerline{\caps ABSTRACT}
In the light of recent  blackhole solutions inspired by string theory, we
review some old statements on field theoretic hair on blackholes. We also
discuss some stability issues. In particular we argue that the two dimensional
string blackhole solution is semi-classically stable while the  naked
singularity  is unstable to tachyon fluctuations. Finally we comment on the
relation
between the linear dilaton theory and the $2d$ blackhole solution.

\endpage

No-hair theorems for blackholes seem to be of two types. The earlier
theorems[\mtw ]   referred to the absence of macroscopic characteristics such
as the higher moments of the mass distribution in a blackhole. In fact these
theorems more or less established that a blackhole is in many respects like an
elementary particle. On the other hand the second set of no-hair theorems
[\beck ],[\adler ]
 seemed to show that blackholes do not have many of the features of elementary
particles either. In particular they cannot have baryon number and more
generally it was believed (though never rigorously established) that
in classical field theoretic blackhole solutions all fields except the static
gravitational, electric, and magnetic, fields go to zero (or unobservable
constant) values outside the horizon. In the case of the physically interesting
spontaneous breakdown of continuous gauge symmetry, arguments were given to the
effect that (at least in the abelian case) the gauge field vanished and the
Higgs field went to a constant value everywhere outside the horizon.

In this paper the assumptions underlying the field theoretic no-hair theorems
are examined. This is motivated in part by the appearance of several recent
articles [\gib ],[\lee ],[\andy ][\al ],[\wit ],[\man], [\dea] (largely
motivated by string theory) which seem to indicate that various types of
(classical) hair (other than the ones permitted by the no-hair theorems) are
possible. In fact it has been conjectured that string theoretic blackholes may
carry an infinite variety of hair [\wit ],[\sch ] ,[\dea], [\ellis]. The way in
which the classical no-hair theorems are voided in these theories is thus of
some interest. In particular the key role played by the dilaton is elucidated.
Next the question of the stability of some of these black hole solutions is
addressed briefly.  In particular the stability  of the Adler-Pearson black
hole for the spontaneously broken abelian Higgs model and the two dimensional
stringy blackhole of [\wit ],[\man ], is discussed.

Let us first review Bekenstein's argument for static blackholes. The horizon of
a blackhole is a null-surface $F(x^i)=0$\foot{$\mu=0,1,2,\ldots d-1\quad
i=1,2,\ldots d-1$ the signature is $(-,+,+,+,...)$ and the definitions of
curvature tensors is as in MTW [\mtw ].} and the normal to it is
$n_{\mu}=\pa_{\mu}F(x^i)$ with the
surface element being $dS_{\mu}=\s n_{\mu}$ with $\s$ being an invariant. If
${\cal L}(\phi_i)$ is a Lagrangian for a set $F$ of fields coupled to gravity,
${\cal M}$ the region outside the horizon with $\pa{\cal M}$ being its
boundary, then the following is easily established from the equations of
motion.

$$\sum_{i\e F'\subseteq F}\int_{\cal M}\sqrt g d^dx \left [\pa{_\mu}\phi_i
{\pa{\cal L}\over\pa\pa_\mu\phi_i}+\phi_i{\pa{\cal L}\over\pa\phi_i}\right]
=\int_{\pa{\cal M}}b_{\mu}dS^{\mu},\eqn\bec$$

where $b^{\mu}=\sum_{i\e F'}\phi_i{\pa{\cal
L}\over\pa\pa_{\mu}\phi_i}$\foot{$\sqrt g=\sqrt |{det g_{\mu\nu}|}$.}. As
Bekenstein points out $g_{ij}$ is a positive definite metric; so it is
possible to show, if $g_{ij}b^ib^j$ is bounded on the horizon, that
 $g_{ij}dS^ib^j=0$ at the horizon. If in addition $b_0$ is
  zero at the horizon then the contribution of the latter to the RHS of \bec is
  zero. At time-like infinity the contribution to the RHS is automatically zero
   and at space-like infinity it is zero if $\phi$ tend to zero at least as
   fast as $1\over r$, $r$ being the coordinate distance from the blackhole. If
   these  conditions are satisfied then the RHS of \bec is zero and if in
   addition the terms in the integrand are positive definite then a no-hair
   theorem is obtained.

 In applications of the above, the constraints on $b$ at the horizon are
obtained by
  the requirement that physical scalars are bounded there. Clearly the trace of
  the stress tensor or any other invariant constructed out of the stress tensor
  should be bounded on the horizon. Let us look first at scalar field theory
  with the Lagrangian

 $${\cal L}=-{1\over 2}(g^{\mu\nu}\pa_{\mu}\phi\pa_{\nu}\phi+V(\phi ))$$

 The stress tensor is

 $$T_{\mu\nu}=\pa_{\mu}\phi\pa_{\nu}\phi-{1\over
 2}g_{\mu\nu}(\pa_{\a}\phi\pa^{\a}\phi+V(\phi ))$$

 $$T_{\mu}^{\mu}={2-d\over 2}\pa_{\a}\phi\pa^{\a}\phi -{d\over 2}V(\phi )$$

 In the static case $\pa_0\phi =0$, and
  $-T_{\mu}^{\mu}={d-2\over 2}g^{ij}\pa_{i}\phi\pa_{j}\phi +{d\over 2}V(\phi
)$, and $g^{ij}$ is positive definite, so for a  (polynomial) potential  whose
highest power in $\phi$ is positive, the finiteness of $T$ implies that $\phi$
is bounded.

 It is important to note here that if a scalar field has no (polynomial)
potential as is the case for the Goldstone boson field, the axion,and the
dilaton then the above argument does not apply. However we may still argue that
the scalar fields are physical quantities which ought to be bounded at the
horizon. In the case of the dilaton for instance, its exponential is the
coupling constant, and one would certainly require the latter to be neither
zero nor infinite at the horizon. We will consider below the case of the axion
and the dilaton blackhole. Solutions having both types of hair have recently
been constructed  [\gib ],[\lee ],[\andy ][\al ],[\wit ],[\man],and it is
instructive to see how precisely they avoid the no-hair theorem.

 The matter action for the axion case is
 $$S_A=\int \sqrt
gd^4x[\pa_{\mu}a\pa^{\mu}a+aF_{\mu\nu}\tilde{F}^{\mu\nu}+...],\eqn\ax$$

 where the ellipses denote $a$ independent terms. From the requirement that the
 trace of the stress tensor is bounded and the positive definiteness of the
spatial metric, we have the result that $g^{ij}\pa_ia\pa_ja$ is bounded at the
horizon.Then assuming that $a$ is a physical scalar which should also be
bounded there, we have from \bec\ the relation
 $$\int d^4x\sqrt g(g^{ij}\pa_ia\pa_ja+aF\tilde{F})=0.$$

 Since the second term is not positive definite we cannot conclude that the
field a vanishes outside the horizon. However if the  either the   electric or
 the magnetic charge of the black hole were to be zero then we would have a
no-hair theorem for the axionic field. This is consistent with the explicit
solution of [\lee ]  which is indeed a dyonic blackhole.

 Let us now consider blackhole solutions coming from the low energy effective
action derived from (classical) string theory [\call ].The latter takes the
following form (keeping just the gauge field and axion terms in addition to the
metric and dilaton) in $d$ dimensional space-time.
 $$S=\int d^dx\sqrt ge^{-2\Phi}  [R+4(\na\Phi )^2
  -{1\over 4}F^2-{1\over 12}H^2-
 {\ov d-26(10)\over 3}+\ldots ]\eqn\effsig$$
$\ov d=d+\d c$ where $\d c$ is the central charge of a compact unitary  cft
which may be tensored with the d target space dimensional sigma model. For a
critical string $\ov d  =26$ (or 10 for the superstring).
Note that the dilaton kinetic term in this action has the wrong sign. For $d>2$
we can do a Weyl transformation $g_{\mu\nu}\rightarrow
e^{4\Phi/(d-2)}g_{\mu\nu}$ to get the action in the canonical form,\foot{Since
the dilaton kinetic energy has the wrong sign the so-called string metric (the
natural metric in the sigma model is probably not the physical metric. In fact
to get the correct dilaton vertex operator around the flat background one
should put $G=e^{4\Phi/(d-2)}\eta\simeq \eta +{4d\over (d-2)}\Phi$ in the sigma
model metric [\sda ].}
$$\eqalign{S=\int d^dx\sqrt g [R-{4\over d-2}(\na\Phi)^2-
{1\over 4}e^{-{4\Phi\over d-2}}F^2&-{1\over 12}e^{-{8\Phi\over
d-2}}H^2\cr&-e^{-{2d\Phi\over d-2}}{\ov d-26(10)\over 3}]\cr}.\eqn\acn$$
Then from \bec\ (taking $\phi_i=\Phi$) one gets

$$\eqalign{\int_{\cal M}\sqrt gd^dx[{-8\over
d-2}g^{ij}\pa_i\Phi\pa_j\Phi+{\Phi\over d-2} (e^{-{4\Phi\over d-2}}F^2+&{2\over
3}e^{-{8\Phi\over d-2}}H^2\cr
+&2de^{-{2d\Phi\over d-2}}{\ov d-26(10)\over 3})]}=0.\eqn\becs$$

Thus  dilatonic hair may exist only if the blackhole carries electromagnetic
and/or axionic charge (or any other type of hair that the stringy blackhole may
allow) or if the corresponding string theory is non-critical. In two dimensions
however a separate discussion is necessary since the above mentioned Weyl
transformation cannot be carried out.
In this case we have only the non-canonical form \effsig\ but the wrong sign of
the dilaton kinetic term does not give rise to a ghost  since there is  no
propagating on mass shell dilaton field in two dimensions. We may treat the
metric which comes from solving the classical equations of motion corresponding
to \effsig\  as the physical metric and ask whether there may be blackhole
solutions with dilatonic hair.Thus from
\effsig\  for $d=2$ \bec\ gives

$$\int_{\cal M}\sqrt gd^2x e^{-2\Phi}\left\{8(1+\Phi )g^{ij}\pa_i\Phi\pa_j\Phi
-2\Phi (R+\ldots )\right\}$$
Even in the absence of gauge or axion fields there is no positive definiteness
so we may have pure dilatonic hair in two dimensions as is of course well known
now from the solutions given in [\wit ],[\man ].

Let us now discuss some questions related to the stability of  blackholes.
First we will consider the blackhole in the abelian Higgs model [\adler ]
.\foot {This is  the closest  to the standard model that we can find in the
literature.}    The model has the Lagrangian
$${\cal L}=-{1\over 4}F^2-|d_{\mu}|^2-V(\phi)\eqn\ab$$
where $V(\phi )=\l (|\phi |^2-\mu^2)^2$ and $d_{\mu}=(\pa_{\mu}-ieA_{\mu
})\phi$. The metric is written in the form
 $$ds^2=-e^{2\a}dt^2+e^{2\b}dr^2+r^2d\O^2\eqn\met$$
 with $e^{2\a}\simeq r-r_H,$ and $e^{2\b}\simeq (r-r_H)^{-1}$ near the horizon
$r=r_H$.
  Following Bekenstein [\beck ] Adler and Pearson [\adler ] choose the gauge
$A_i=0$, $\pa_0 A_0=0$ and $\phi$ real, for the static case.\foot{Actually it
is not clear whether this is a valid choice if one wishes to preserve the
asymptotic conditions at spatial infinity which are necessary to put the
integral over that surface to zero in the RHS of \bec .} Using the boundedness
of the stress tensor in an orthonormal frame to get boundary conditions on the
horizon it is argued in [\adler ] that $A_0$ is zero everywhere outside the
  horizon. In the symmetric case ($\mu^2<0$,$\phi\rightarrow 0$ asymptotically)
 on the other hand one gets $\phi =0$ outside while $A_0\ne 0$. As in the case
of flat space for the $\mu^2\ge 0$ case it is possible to show that the $\phi
=\mu$
 $A_0=0$ (outside the blackhole) solution is stable under  scalar field
fluctuations. The field equations are

$$G_{\mu\nu}=T_{\mu\nu},\quad\na^{\mu}F_{\mu\nu}
=j_{\nu},\quad\na^2_A\phi=V'(\phi ),$$
 where $$T_{\mu\nu}={1\over 2}F_{\mu\l}F_{\nu}^{\l}-{1\over
8}g_{\mu\nu}F^2+d_{(\mu}d_{\nu )}^*-{1\over 2}g_{\mu\nu}(d_{\l}d^{\l *}-V(\phi
0))$$
 and

$$j_{\l}=-ie(\phi\pa_{\l}\phi^{*}-\phi^{*}\pa_{\l}\phi)+2e^2A_{\l}\phi\phi^*.$$
 We need to look at fluctuations around the background $A=0$, $\phi =\mu$, with
the metric given by \met . It is then  seen from the above equations that for
real fluctuations of $\phi$ the fluctuations of the metric and the gauge field
are such that $\d g\simeq\d A\simeq O((\d\phi )^2)$, Thus in analysing the
stability, at least under real fluctuations of $\phi$, we are not constrained
to  take into account the fluctuations of $A$ and $g$. This simplifies the
problem considerably. The fluctuations must also satisfy the condition at the
horizon coming from the requirement that physical scalars should not blow up
there. This implies that $\d\phi$ should vanish at least as fast as
$(r-r_H)^{1\over 2}$ as $r\rightarrow r_H$. In addition of course the
fluctuations must satisfy
 the boundary condition at infinity, namely fall off at least as fast as
 ${1\over r}$. It is sufficient to consider s-wave fluctuations since the
angular momentum barrier contributes a repulsive potential. Thus putting
$\d\phi =e^{\o t}f(r)$ and using the background values of the metric gauge and
scalar fields we have the following equations for the linearized fluctuations;

 $$ -\o^2f= -{e^{\a-\b}\over
r^2}\pa_r(r^2e^{\a-\b}\pa_rf(r))+4e^{2\a}h\mu^2f(r).$$
 Changing variable to $x$ such that  ${dx\over dr}={e^{\b-\a}\over r^2}$ one
gets after multiplying by $r(x)^4$, a Schrodinger operator on the RHS. Given
the boundary conditions on the fluctuations  and the positive definiteness of
the  potential it is easily seen that this operator is positive definite. Thus
there are no bound states and the blackhole is stable under such perturbations.
On the other hand (as  in   flat space)  the $\phi\rightarrow 0$, $A\ne 0$,
solution is unstable.

 Similar arguments can be used to show that the 2d string theoretic blackholes
are stable under tachyonic perturbations. Since the tachyon is the only
physical propagating field in two dimensional string theory this is the only
perturbation we need to consider. The relevant equations (for large black holes
so that we can use the leading order equations in $\a '$ outside the horizon)
 are

 $$\eqalign {0 & = R_{\mu\nu}+2\na_{\mu}\na_{\nu}\phi-\na_{\mu}
T\na_{\nu}T+\ldots\cr
0 & = -R+4(\na\Phi )^2-4\na^2\Phi +(\na T)^2-2T^2
- 8+\ldots\cr
0 & = -2\na^2T+4\na\Phi \na T+V'(T)+\ldots\cr}\eqn\bet$$
where the ellipses indicate higher derivative terms, higher powers of the
tachyon, etc. As shown by Mandal et al [\man ]  for $T=0$ the general solution
of this set of equations may be put in the form ($X^{\mu}=(t,\phi )$)
$$\eqalign{ds^2 &\simeq -(1-ae^{-2\sqrt 2\phi})dt^2+
(1-ae^{-2\sqrt 2\phi})^{-1}d\phi^2\cr }\eqn\metric$$

 with $\Phi =\sqrt 2\phi$ and $a$ is the  black hole mass up to a positive
constant. This is of course the low energy metric obtained from Witten's
${SL(2,R)\over U(1)}$ theory. From \bet\  it is easy to see that this solution
is stable under linearized fluctuations. Firstly from the first two equations
 of \bet  we see that $T$ affects the dilaton and graviton only at second order
so that at linear order we can ignore the metric and dilaton fluctuations. In
terms of the physical tachyon $S=e^{-\Phi }T$ the tachyon equation becomes
 $$\na^2S-{R\over 4}S=0$$
 Now for the above metric $R=8ae^{-2\sqrt 2\phi}$ so for positive $a$ (the
blackhole case) we have from the same type of argument as  we made earlier,
 the result that the solution \metric  is stable. On the other hand for $a<0$,
the case of a naked singularity, the solution is unstable.

 To analyze the quantum stability we need the Euclidean effective action for
the tachyon fluctuations. Since the two-dimensional blackhole is a particular
background
 for two-dimensional closed string field theory we may obtain it from the
 background independent formulation [\ramy ] of the Das-Jevicki  theory [\dj ]
since the latter at least in the flat space case seems to be a string field
theory [\jev ]. Hence it is reasonable to expect that the usual arguments \foot
{see for instance [\cole ].} for analyzing quantum stability under small
fluctuations are valid when applied to the background independent form of the
Das-Jevicki theory.
 The quadratic (in $S$) piece of the tachyon action is (from equation (4) of
 [\ramy ])
 $$\int d^2x\sqrt g(\na S\cdot\na S+{1\over 4}RS^2)\eqn\quad$$

 and by the usual argument semi-classical (meta)stability depends on the
spectrum of the euclidean signature operator
 $$-\na_E^2+R.$$
If the above operator has no negative eigenvalues then the background is
semi-classically stable at least under tachyon fluctuations. Since $-\na^2_E$
is a positive definite operator we have the same result as in the classical
case: stability for the blackhole (since $R>0$) and instability for the naked
singularity.
 It should be noted that the quadratic tachyon action above is equivalent to
the quadratic part of the usual tachyon action for the sigma model tachyon
 $$\int d^2x e^{-2\Phi}\sqrt g((\na T)^2 -2T^2 )$$
 where $T=e^{\Phi}S$ provided we use the lowest order (in $\a '$) beta function
equation for the dilaton (second equation in \bet ). In this latter form the
quadratic action for tachyons is not positive definite and indeed has  negative
eigenvalues.However the corresponding eigenfunctions  $S$ are not square
integrable; For instance the eigenfunction $T={1/\sqrt V}$ where $V=\int \sqrt
g$
 corresponds to $S=\cosh r/\sqrt V$ for $S$. Our point of view is that S is the
physical tachyon and it is also the field which comes naturally  in the
Das-Jevicki action and the string field theory functional measure should be
derived from the metric $\parallel \d S\parallel^2=\int\sqrt g\d S^2$. In other
words fluctuations of the sigma model tachyon $T$ should be normalized  using
the metric $||\d T||=\int\sqrt g\d T^2||$. With this interpretation of the
string field theory the blackhole solution is semi-classically stable.\foot{The
stability of the two dimensional blackhole under tachyonic fluctuations has
also been established by a somewhat different argument in [\ss ] }On the other
hand  the naked singularity is unstable.\foot{Of course we are assuming here
that the existence of negative eigenvalues in the Euclidean space operator
governing quadratic fluctuations makes the correspoding Minkowski space
configuration  unstable.}

  Is there a possibility that some stringy effect destabilizes the blackhole.
 The  argument in [\ramy ] on the existence of the $RS^2$ term in the effective
action relied in part on  lowest order in $\a '$ expressions. In principal even
if  the Das-Jevicki theory contains all tachyon-dilaton terms which are at
least linear in the tachyon (this may well be the case since the theory
corresponds to a string moving in a flat background with a linear dilaton ) it
certainly cannot say anything about curvature tachyon terms. There is  evidence
(from stringy beta-function calculations [\wat ]) also
 for a $R^2S$ term in the tachyon beta function. This would then mean that at
non-leading order in $\a '$ the sigma model interpretation of the
$SL(2,R)/U(1)$ blackhole necessarily contains a non-zero tachyon background
[\dea ]. Indeed general arguments [\dea ] seem to indicate that the string
theoretic
blackhole excites all higher spin modes as well.

The above argument of course does not mean that the blackhole is stable.First
of all there is the integral over dilaton and metric fluctuations. As usual in
Euclidean gravity this may be problematic.\foot{In conformal gauge the
(non-diagonal) metric for  kinetic terms of the dilaton and Liouville mode is
not positive definite.} In addition one  knows that it must decay by Hawking
radiation. The standard interpretation of the Euclidean blackhole solution is
that it corresponds to a blackhole in equilibrium with a heat bath [\hh] at the
Hawking temperature (in this case $T_H=2\sqrt 2/4\pi$). This interpretation
however  seems to require that the vacuum solution (or metric dilaton solution
in the $2d$  case) remains valid even in a heat bath. Also being an equilibrium
picture it does not shed light on the actual decay of the hole. In particular
the nature of  the end point  is not resolved.

In the $2d$ case Witten [\wit ] has conjectured that the end point of decay is
the linear dilaton theory. In fact it is easy to see that the Euclidean action
is a monotonically decreasing function of the blackhole mass (provided the
infrared divergence is cutoff). Substituting the solution
\metric and $\Phi=2\sqrt 2\phi$ in the Euclidean effective action (with $T=0$)
 $$S_E=-\int e^{-2\Phi} (R+4(\na\Phi )^2+8) $$
 and noting that for this euclidean (cigar) solution  the coordinate range for
$\phi$ is $[-\infty,-\ln a/2\sqrt 2]$ we have

 $$S_E^{BH} =-16\int_{-\infty}^{-\ln a/2\sqrt 2}e^{-2\sqrt 2 \phi}$$

 Cutting off the lower end of the integral to define it we see that action
decreases with the blackhole mass  $(M\simeq a)$ with the lowest action
corresponding to the $a=0$ case i.e. the linear dilaton theory. Now one might
ask why the system doesn't simply slide down from non-zero $a$ to zero. As
pointed out by Shenker and Seiberg [\ss ] the reason is that the measure on the
space of small  metric deformations which change $a$, $\d g={\pa g\over\pa a}$,
is not defined . The corresponding metric is given by
 $$||\d g||^2=\int \sqrt g(g^{\a\g}g^{\b\d}+g^{\a\d}g^{\b\g})\d g_{\a\b}\d
g_{\g\d}\simeq\int_{-\infty}^{-\ln a/2\sqrt 2}d\phi d\tau
 {e^{4\sqrt 2\phi}\over(1-ae^{2\sqrt 2\phi})^2} ,$$

and clearly diverges at the tip of the cigar.\foot{ It should be noted however
that this is not
the case if the black hole metric is defined in the (Euclidean conformal gauge
$ds^2={1\over 2z\ov z+a}dzd\ov z$. In that case one gets
$||\d g||^2=4\int {dzd\ov z\over (2z\ov z+a)^3}$ which is finite for $a\ne 0$.
It does blow up
for $a=0$ though, so that the linear dilaton theory is still (perturbatively)
isolated from the
space of blackhole configurations.}
Thus there are no (square
integrable) metric perturbations which can change the mass of the blackhole so
that the black hole cannot decay perturbatively. The effect of this is to erect
a sort of potential barrier (or perturbative superselection rule? [\ss ])
between different values of $a$ and  hawking decay of the blackhole must take
place as a tunneling process.The quantum-mechanical wave function given by the
path integral must be a superposition of  undecayed and  decayed components and
its path integral representation clearly must include an instruction to
integrate over all classical solutions, i.e. in this case all values of
$a$.Thus in
order to discuss the decay of one blackhole of a given mass in this formalism
one seems to require a solution to the problem of the collapse of the wave
function to one classical solution. At least in this two dimensional model of
quantum gravity one knows the whole space of classical solutions [\man ] and so
 it is perhaps a good laboratory to study these questions.

\chapter{\bf Acknowledgements} I would like to thank Ramy Brustein, Gary
Horowitz, and Joe Polchinski,  for discussions. This work  is partially
supported by Department of Energy contract No.
DE-AC02-86ER40253.

\refout

 \end